# On the EPR paradox and local causality principle

*Zbigniew Zimpel*

Signify Intellectual Property, High Tech Campus 7, 5656 AE Eindhoven, The Netherlands

E-mail: zbz900@mail.usask.ca

**Abstract**

Formulas for calculating the joint probability of outcomes of measurements performed on mutually non-interacting component systems of a combined system prepared in an entangled state are presented. The formulas are based on non-relativistic quantum mechanics. An interpretation of the joint probability which conforms to the principle of local causality is proposed. The joint probability of outcomes of spin measurements performed on the well-known system of two spatially separated spin-1/2 particles prepared initially in the singlet state is used to illustrate the proposed interpretation of theoretical and experimental results.

## 1. Introduction

In their famous 1935 paper [6], Einstein, Podolsky, and Rosen (EPR) described a Gedankenexperiment which in their opinion led to the conclusion that quantum mechanics may be incomplete. The EPR paper has attracted much attention among both physicists and philosophers of science who for more than 85 years have been trying to prove either that quantum mechanics is indeed incomplete and needs additional information to describe physical reality or the opposite, that quantum mechanics as such is complete. An overview of various aspects of the EPR paradox is available in [9].

The objective of the present article is to shed some light on a possible flaw in the reasoning presented in the EPR paper and to provide answers to some questions raised or inspired by its authors based on non-relativistic quantum mechanics and local realism. The key for obtaining a correct solution to the problem exposed in the EPR paper is the proper interpretation of the joint probability of outcomes of multiple measurements performed on mutually non-interacting component systems of a combined system which was in an entangled state before the component systems ceased to interact with each other.

Suppose that there are two component systems I and II of a combined system I + II. Let $\{|\varphi_i\rangle: i = 1, \ldots, M\}$ and $\{|\psi_j\rangle: j = 1, \ldots, N\}$ be two sets of state vectors forming orthonormal bases of finite-dimensional Hilbert spaces $\mathcal{H}_1$ and $\mathcal{H}_2$ associated with system I and II, respectively. The tensor product $\mathcal{H}_1 \otimes \mathcal{H}_2$ of Hilbert spaces $\mathcal{H}_1$ and $\mathcal{H}_2$ is the linear span of tensor products $|\varphi_i\rangle \otimes |\psi_j\rangle$ of basis vectors in $\mathcal{H}_1$ and $\mathcal{H}_2$. $\mathcal{H}_1 \otimes \mathcal{H}_2$ is a Hilbert space associated with the combined system I + II. The products $|\varphi_i\rangle \otimes |\psi_j\rangle$ form an orthonormal basis of $\mathcal{H}_1 \otimes \mathcal{H}_2$. Hereinafter, a tensor product $|\varphi\rangle \otimes |\psi\rangle$ of two vectors $|\varphi\rangle$ in $\mathcal{H}_1$ and $|\psi\rangle$ in $\mathcal{H}_2$ will be abbreviated as $|\varphi \otimes \psi\rangle$.

Every state of the combined system I + II is represented by a state vector $|\Psi\rangle$ in $\mathcal{H}_1 \otimes \mathcal{H}_2$ and can be written as a normalized linear combination of basis vectors $|\varphi_i \otimes \psi_j\rangle$:

$$|\Psi\rangle = \sum_{i,j} c_{ij} |\varphi_i \otimes \psi_j\rangle. \tag{1}$$

A state $|\Psi\rangle$ of the combined system I + II is not entangled if there exists a pair of states $|\varphi\rangle$ in $\mathcal{H}_1$ and $|\psi\rangle$ in $\mathcal{H}_2$ such that $|\Psi\rangle = |\varphi \otimes \psi\rangle$. Otherwise, the state is called an entangled state.

Let $A$ denote an arbitrary observable pertaining to system I, i.e., a self-adjoint operator on $\mathcal{H}_1$, and let $|\alpha_1\rangle, |\alpha_2\rangle, \ldots, |\alpha_M\rangle$ denote eigenvectors of $A$ corresponding to eigenvalues $a_1, a_2, \ldots, a_M$. Then $A \otimes I_2$, where $I_2$ denotes the identity operator on $\mathcal{H}_2$, is an observable pertaining to the combined system I + II, corresponding to $A$. Operators $A$ and $A \otimes I_2$ have identical eigenvalues. The linear span $\mathcal{H}_a = Span\{|\alpha_i\rangle : a_i = a\}$ is the eigenspace of $A$ associated with an eigenvalue $a$, and $\mathcal{H}_a \otimes \mathcal{H}_2$ is the eigenspace of $A \otimes I_2$ associated with the eigenvalue $a$. Further, let $B$ denote an arbitrary observable pertaining to system II, i.e., a self-adjoint operator on $\mathcal{H}_2$, and let $|\beta_1\rangle, |\beta_2\rangle, \ldots, |\beta_N\rangle$ denote eigenvectors of $B$ corresponding to eigenvalues $b_1, b_2, \ldots, b_N$. Then $I_1 \otimes B$, where $I_1$ denotes the identity operator on $\mathcal{H}_1$, is an observable pertaining to the combined system I + II, corresponding to $B$. Operators $B$ and $I_1 \otimes B$ have identical eigenvalues. The linear span $\mathcal{H}_b = Span\{|\beta_j\rangle : b_j = b\}$ is the eigenspace of $B$ associated with an eigenvalue $b$, and $\mathcal{H}_1 \otimes \mathcal{H}_b$ is the eigenspace of $I_1 \otimes B$ associated with the eigenvalue $b$.

It is assumed throughout this article that the Hamiltonian of each component system before and after the measurement is the identity operator on the respective Hilbert space multiplied by an energy constant. Hence, every state vector of the component system is an eigenvector of the Hamiltonian and corresponds to the same eigenvalue equal to the energy constant. This assumption guarantees that every observable pertaining to the component system commutes with the Hamiltonian of the component system and therefore can be measured, and that after a collapse due a measurement performed on a component system, the combined system remains in the state resulting from the collapse.

The structure of the present article is as follows. In Sec. 2, the EPR paper is briefly summarized using the notation and assumptions introduced above. The formula for the joint probability of outcomes of measurements performed on mutually non-interacting component systems of a combined system, when the combined system was initially in a state $|\Psi\rangle$, is presented in Sec. 3. In Sec. 4, the postulate of the local collapse of an initial entangled state of the combined system comprising two component systems initially interacting with each other to a non-entangled state when the component systems cease to interact, is formulated. Using this postulate, the joint probability derived in Sec. 3 is shown to conform with the principle of local causality. The results of this article are illustrated in Sec. 5, where they are applied to the system of two spin-1/2 particles in the singlet state.

**2. EPR paradox**

Suppose that the two component systems are initially allowed to interact, but after some time there is no longer interaction between them. According to the EPR paper, after the component systems cease to interact, the combined system is still in an initial entangled state $|\Psi\rangle$ which is an eigenstate of the interaction Hamiltonian. Assume that a measurement of an observable $A$ gives a non-degenerate eigenvalue $a_i$. According to the EPR paper, after the measurement the system collapses from the initial state $|\Psi\rangle$ to a state defined by the vector

$$|\alpha_i\rangle \otimes \sum_j \langle \alpha_i \otimes \psi_j | \Psi \rangle |\psi_j\rangle = |\alpha_i\rangle \otimes |\xi_i\rangle, \qquad (2)$$

where

$$|\xi_i\rangle = \sum_j \langle \alpha_i \otimes \psi_j | \Psi \rangle |\psi_j\rangle \qquad (3)$$

defines the state vector of system II after the measurement of $A$. $|\xi_i\rangle$ must be normalized to represent a state vector. If, instead of measuring the observable $A$, one chooses to measure another observable, say $A'$, having eigenvectors $|\alpha'_1\rangle, |\alpha'_2\rangle, \ldots, |\alpha'_M\rangle$ in $\mathcal{H}_1$ associated with eigenvalues $a'_1, a'_2, \ldots, a'_M$, and the measurement gives a non-degenerate eigenvalue $a'_k$, then after the measurement the system collapses from $|\Psi\rangle$ to a state defined by the vector

$$|\alpha'_k\rangle \otimes \sum_j \langle \alpha'_k \otimes \psi_j | \Psi \rangle |\psi_j\rangle = |\alpha'_k\rangle \otimes |\xi'_k\rangle, \tag{4}$$

where

$$|\xi'_k\rangle = \sum_j \langle \alpha'_k \otimes \psi_j | \Psi \rangle |\psi_j\rangle \tag{5}$$

defines the state vector of system II after the measurement of $A'$. $|\xi'_k\rangle$ must be normalized to represent a state vector.

Based on Eq. (2) and (4), the authors of the EPR paper state: "*We see therefore that, as a consequence of two different measurements performed upon the first system, the second system may be left in states with two different wave functions. On the other hand, since at the time of measurement the two systems no longer interact, no real change can take place in the second system in consequence of anything that may be done to the first system.*" Hence, on the one hand, the states $|\xi_i\rangle$ and $|\xi'_k\rangle$ should correspond to the same physical reality and on the other hand, "*it is in general possible for* [the states $|\xi_i\rangle$ and $|\xi'_k\rangle$ to be eigenvectors] *of two non-commuting operators, corresponding to physical quantities*" and therefore, they may correspond to different physical realities. In a nutshell, this is the paradox described in the EPR paper.

Einstein realized that accepting a global collapse of an entangled state resulting from a locally performed measurement resolves the EPR paradox as formulated in the EPR paper, but he was seriously concerned about the implications of accepting such non-local effect of measurements performed locally on mutually non-interacting component systems, which he described as a "*spooky action at a distance*" [7].

## 3. Joint probability of outcomes of multiple measurements

In this section, formulas for the joint probability of outcomes of measurements performed on mutually non-interacting component systems of a combined system, when the combined system was initially in a state $|\Psi\rangle$, are derived using non-relativistic quantum mechanics as presented for example in [12]. Consider a first measurement of an observable $A$ performed on system I. The probability that the measurement of $A$ gives an eigenvalue $a$ is

$$\mathbb{P}(A, a|\Psi) = \left\|P_{A,a} \otimes I_2 \Psi\right\|^2 = \langle P_{A,a} \otimes I_2 \Psi | \Psi \rangle = \sum_{\{i: a_i = a\}} \sum_j |\langle \alpha_i \otimes \psi_j | \Psi \rangle|^2. \tag{6}$$

Here $P_{A,a}$ is the orthogonal projection operator from $\mathcal{H}_1$ onto $\mathcal{H}_a$ and $P_{A,a} \otimes I_2$ is the orthogonal projection operator from $\mathcal{H}_1 \otimes \mathcal{H}_2$ onto $\mathcal{H}_a \otimes \mathcal{H}_2$. It is assumed that due to the measurement, the combined system collapses from state $|\Psi\rangle$ to a new state defined by the orthogonal projection of $|\Psi\rangle$ onto $\mathcal{H}_a \otimes \mathcal{H}_2$:

$$P_{A,a} \otimes I_2 |\Psi\rangle = |P_{A,a} \otimes I_2 \Psi\rangle = \sum_{\{i: a_i = a\}} \sum_j \langle \alpha_i \otimes \psi_j | \Psi \rangle |\alpha_i \otimes \psi_j\rangle. \tag{7}$$

$|P_{A,a} \otimes I_2 \Psi\rangle$ must be normalized to represent a state vector. If the eigenvalue $a$ is non-degenerate, then $|P_{A,a} \otimes I_2 \Psi\rangle$ is not entangled. Otherwise, it may be entangled.

Suppose now that after the first measurement of the observable $A$ which gave the eigenvalue $a$, a second measurement of an observable $B$ is performed and gives an eigenvalue $b$. The joint probability of the two outcomes is

$$\mathbb{P}(B, b|A, a|\Psi) = \left\|(I_1 \otimes P_{B,b})(P_{A,a} \otimes I_2)\Psi\right\|^2 = \left\|P_{A,a} \otimes P_{B,b} \Psi\right\|^2 = \langle P_{A,a} \otimes P_{B,b} \Psi | \Psi \rangle$$

$$= \sum_{\{i: a_i = a\}} \sum_{\{j: b_j = b\}} |\langle \alpha_i \otimes \beta_j | \Psi \rangle|^2. \tag{8}$$

Here $P_{B,b}$ is the orthogonal projection operator from $\mathcal{H}_2$ onto $\mathcal{H}_b$, $I_1 \otimes P_{B,b}$ is the orthogonal projection operator from $\mathcal{H}_1 \otimes \mathcal{H}_2$ onto $\mathcal{H}_1 \otimes \mathcal{H}_b$, and $P_{A,a} \otimes P_{B,b}$ is the orthogonal projection operator from $\mathcal{H}_1 \otimes \mathcal{H}_2$ onto $\mathcal{H}_a \otimes \mathcal{H}_b$. The following properties of the projection operators $P_{A,a} \otimes I_2$ and $I_1 \otimes P_{B,b}$ were used to calculate the probabilities in Eq. (6) and (8): the two operators (i) commute with each other; (ii) are self-adjoint; and (iii) are idempotent.

Since $A \otimes I_2$ and $I_1 \otimes B$ commute with each other, the joint probability (8) does not depend on the sequence in which the two measurements are performed. It is also identical with the joint probability of outcomes when the two measurements are performed simultaneously. Thus, one can denote the joint probability without indicating the order in which the two measurements are performed:

$$\mathbb{P}(A, a, B, b|\Psi) = \mathbb{P}(B, b|A, a|\Psi) = \mathbb{P}(A, a|B, b|\Psi). \tag{9}$$

If $|\Psi\rangle = |\varphi \otimes \psi\rangle$ for some state vectors $|\varphi\rangle$ in $\mathcal{H}_1$ and $|\psi\rangle$ in $\mathcal{H}_2$, the joint probability $\mathbb{P}(A, a, B, b|\Psi)$ can be factorized as follows:

$$\mathbb{P}(A, a, B, b|\varphi \otimes \psi) = \mathbb{P}(A, a|\varphi \otimes \psi)\mathbb{P}(B, b|\varphi \otimes \psi) = \|P_{A,a}\varphi\|^2 \|P_{B,b}\psi\|^2. \tag{10}$$

Using the joint probability (8) one can calculate the quantum correlation function defined as

$$C_q = \sum_{a,b} ab \mathbb{P}(B, b|A, a|\Psi) = \sum_{a,b} ab \langle P_{A,a} \otimes P_{B,b} \Psi|\Psi\rangle = \langle A \otimes B \Psi|\Psi\rangle. \tag{11}$$

Using Eq. (10), it is straightforward to verify that if $|\Psi\rangle$ is not entangled the quantum correlation can be factorized.

The following result generalizes Eq. (8) to deal with the case of multiple measurements. Let $R_1, R_2, \ldots, R_m$ be a first sequence of observables pertaining to system I and let $S_1, S_2, \ldots, S_n$ be a second sequence of observables pertaining to system II. Let $Q_1, Q_2, \ldots, Q_{m+n}$ denote a sequence of observables pertaining to the combined system I + II consisting of two subsequences:

$R_1 \otimes I_2, R_2 \otimes I_2, \ldots, R_m \otimes I_2$ and $I_1 \otimes S_1, I_1 \otimes S_2, \ldots, I_1 \otimes S_n$. Assume that the measurements of observables $Q_1, Q_2, \ldots, Q_{m+n}$ gave eigenvalues $r_1, r_2, \ldots, r_m$ of observables $R_1, R_2, \ldots, R_m$ and eigenvalues $s_1, s_2, \ldots, s_n$ of observables $S_1, S_2, \ldots, S_n$. Then the joint probability that the first measurement of $Q_1$ gave a value $q_1$, the second measurement of $Q_2$ gave a value $q_2$, …, and the $(m+n)$-th measurement of $Q_{m+n}$ gave a value $q_{m+n}$ is

$$\mathbb{P}(Q_{m+n}, q_{m+n}| \ldots |Q_1, q_1|\Psi) = \left\|\left(P_{R_m, r_m} \ldots P_{R_1, r_1}\right) \otimes \left(P_{S_n, s_n} \ldots P_{S_1, s_1}\right)\Psi\right\|^2. \tag{12}$$

Indeed, since the tensor product $\left(P_{R_m, r_m} \ldots P_{R_1, r_1}\right) \otimes \left(P_{S_n, s_n} \ldots P_{S_1, s_1}\right)$ is identical to the product $\left(P_{R_m, r_m} \otimes I_2\right) \ldots \left(P_{R_1, r_1} \otimes I_2\right)\left(I_1 \otimes P_{S_n, s_n}\right) \ldots \left(I_1 \otimes P_{S_1, s_1}\right)$ and since operators $P_{R_k, r_k} \otimes I_2$ and $I_1 \otimes P_{S_l, s_l}$ commute with each other for every $1 \leq k \leq m$ and $1 \leq l \leq n$, the factors $P_{R_k, r_k} \otimes I_2$ can be rearranged with respect to the factors $I_1 \otimes P_{S_l, s_l}$ to obtain the arrangement identical with the arrangement of operators $P_{R_k, r_k} \otimes I_2$ and $I_1 \otimes P_{S_l, s_l}$ in the product $P_{Q_{m+n} q_{m+n}} \ldots P_{Q_1 q_1}$ defining the joint probability (12).

## 4. The principle of local causality

Intuition says that measurements performed on one system should not influence the outcomes of measurements performed on the other system when the two systems do not interact with each other. The outcome of a measurement performed on either system should depend only on the measured observable and the state of the system on which the measurement is performed. His intuition led Einstein to question the possibility of a "*spooky action at a distance*" due to the collapse of an entangled state of two physically isolated systems when an observable pertaining to one of these systems is measured. This issue revealed in the EPR paper relates to the local causality principle and was addressed by Bell and many others [2, 3, 13].

Bell's locality (factorizability) condition is the assumption of Bell's theorem [13]. Bell considered this condition a consequence of local causality: if outcomes of two measurements performed locally on mutually non-interacting systems are locally explicable, i.e., are not in a cause-effect relation, then the joint probability of their outcomes can be factorized, i.e., the joint probability of their outcomes can be expressed as a product of the probability of an outcome of the first measurement and the probability of an outcome of the second measurement. When the joint probability of outcomes of two measurements can be factorized, the outcomes are called uncorrelated, statistically independent, or just independent. However, the contraposition of Bell's assumption implies that if the joint probability cannot be factorized, which is equivalent to saying that the outcomes of these measurements are correlated, then they are in a cause-effect relation. This disagrees with the principle "correlation does not imply causation", which is well recognized and widely accepted is statistics and classical probability calculus [11]. A more in-depth discussion of Bell's locality condition can be found in [3, 13].

There are two possible reasons justifying a correlation between outcomes of measurements of two observables $A$ and $B$ performed on two component systems: the measurement of $A$ performed on system I has an effect on the outcome of the subsequent measurement of $B$ performed on system II and/or the outcomes of measurements of $A$ and $B$ are related by a common cause [11]. Since systems I and II do not interact with each other and each measurement apparatus interacts only with the system on which the measurement is performed, the first possibility should be rejected.

When the component systems I and II interact with each other, the state vector $|\Psi\rangle$ of the combined system I + II is entangled and no states can be assigned to the component systems. The entangled state $|\Psi\rangle$ is an eigenvector of the Hamiltonian of the interaction between the component systems. It is postulated here that when the two component systems cease to interact with each other (prior to any measurement), the initially entangled state $|\Psi\rangle$ locally collapses to a non-entangled state defined by a tensor product $|\xi\otimes\zeta\rangle$ of a pair of state vectors $|\xi\rangle$ and $|\zeta\rangle$ in the respective Hilbert spaces $\mathcal{H}_1$ and $\mathcal{H}_2$. To this end, it will be now shown that

$$|\Psi\rangle = MN \int |U_1\varphi \otimes U_2\psi\rangle\langle U_1\varphi \otimes U_2\psi|\Psi\rangle dU_1 dU_2 \tag{13}$$

for any pair of state vectors $|\varphi\rangle$ in $\mathcal{H}_1$ and $|\psi\rangle$ in $\mathcal{H}_2$, where $|U_1\varphi\rangle \equiv U_1|\varphi\rangle$, $|U_2\psi\rangle \equiv U_2|\psi\rangle$ and the integration is carried out over all matrices $U_1$ of the unitary group $U_1(M)$ with respect to the normalized Haar measure $dU_1$ and over all matrices $U_2$ of the unitary group $U_2(N)$ with respect to the normalized Haar measure $dU_2$. Indeed, for every vector $|\Phi\rangle$ in $\mathcal{H}_1\otimes\mathcal{H}_2$

$$\int \langle\Phi|U_1\varphi\otimes U_2\psi\rangle\langle U_1\varphi\otimes U_2\psi|\Psi\rangle dU_1 dU_2 = \int \langle\Phi|U_1\varphi_i\otimes U_2\psi_j\rangle\langle U_1\varphi_i\otimes U_2\psi_j|\Psi\rangle dU_1 dU_2 \tag{14}$$

since there exist $V_1 \in U_1(M)$ and $V_2 \in U_2(N)$ such that $|\varphi\rangle = V_1|\varphi_i\rangle$ and $|\psi\rangle = V_2|\psi_j\rangle$, and the Haar measures $dU_1$ and $dU_2$ are right-invariant, i.e., $\int f(U_i)dU_i = \int f(U_i V_i)dU_i$ for every $dU_i$-integrable function $f: U_i \to \mathbb{C}$ and every matrix $V_i \in U_i$. Then

$$\int \langle\Phi|U_1\varphi_i\otimes U_2\psi_j\rangle\langle U_1\varphi_i\otimes U_2\psi_j|\Psi\rangle dU_1 dU_2$$

$$= \sum_{k,l,k',l'} \langle\Phi|\varphi_k\otimes\psi_l\rangle\langle\varphi_{k'}\otimes\psi_{l'}|\Psi\rangle \int \langle\varphi_k\otimes\psi_l|U_1\varphi_i\otimes U_2\psi_j\rangle\langle U_1\varphi_i\otimes U_2\psi_j|\varphi_{k'}\otimes\psi_{l'}\rangle dU_1 dU_2$$

$$= \sum_{k,l,k',l'} \langle\Phi|\varphi_k\otimes\psi_l\rangle\langle\varphi_{k'}\otimes\psi_{l'}|\Psi\rangle \int \langle\varphi_k|U_1\varphi_i\rangle\langle U_1\varphi_i|\varphi_{k'}\rangle dU_1 \int \langle\psi_l|U_2\psi_j\rangle\langle U_2\psi_j|\psi_{l'}\rangle dU_2$$

$$= \frac{1}{MN} \sum_{k,l} \langle\Phi|\varphi_k\otimes\psi_l\rangle\langle\varphi_k\otimes\psi_l|\Psi\rangle = \frac{1}{MN}\langle\Phi|\Psi\rangle \tag{15}$$

since using the invariant integration method described in [1] one can see that

$$\int \langle \varphi_k | U_1 \varphi_i \rangle \langle U_1 \varphi_i | \varphi_{k'} \rangle dU_1 = \frac{\delta_{kk'}}{M},$$
$$\int \langle \psi_l | U_2 \psi_j \rangle \langle U_2 \psi_j | \psi_{l'} \rangle dU_2 = \frac{\delta_{ll'}}{N}. \tag{16}$$

Equation (14) and (15) proves that Eq. (13) holds.

The probability amplitude $\langle \alpha_i \otimes \beta_j | \Psi \rangle$ in Eq. (8) can now be written as

$$\langle \alpha_i \otimes \beta_j | \Psi \rangle = MN \int \langle \alpha_i | U_1 \varphi \rangle \langle \beta_j | U_2 \psi \rangle \langle U_1 \varphi \otimes U_2 \psi | \Psi \rangle dU_1 dU_2. \tag{17}$$

Since for every state vector $|\xi\rangle$ and $|\zeta\rangle$ there exist $U_1 \in U_1(M)$ and $U_2 \in U_2(N)$ such that $|\xi\rangle = |U_1 \varphi\rangle$ and $|\zeta\rangle = |U_2 \psi\rangle$, the superposition of state vectors in Eq. (13) and the integral of probability amplitudes in Eq. (17) includes all state vectors in the respective Hilbert spaces $\mathcal{H}_1$ and $\mathcal{H}_2$. The factor $MN$ in Eq. (13) and (17) can be used to renormalize the Haar measures $dU_1$ and $dU_2$.

The information about the initially entangled state (common cause) is carried over by the probability amplitudes $\langle U_1 \varphi \otimes U_2 \psi | \Psi \rangle$ of the local collapse of $|\Psi\rangle$ to $|U_1 \varphi \otimes U_2 \psi\rangle$. However, it is impossible to determine the actual state vectors $|U_1 \varphi\rangle$ and $|U_2 \psi\rangle$ after the local collapse based on the state vector $|\Psi\rangle$. Therefore, classical probability calculus fails to explain the correlation between the outcomes of measurements performed on component systems after their local collapse. This situation resembles the effect encountered in the double-slit experiment. As Feynman pointed out [8], "*When an event can occur in several alternative ways, the probability amplitude for the event is the sum of the probability amplitudes for each way considered separately.*"

Now it is straightforward to explain why the outcomes of measurements of observables $A$ and $B$ are correlated. To this end it is helpful to rewrite the joint probability (8) as follows:

$$\mathbb{P}(B,b|A,a|\Psi) = \sum_{\{i:a_i=a\}} \sum_{\{j:b_j=b\}} |\langle \alpha_i \otimes \beta_j|\Psi\rangle|^2$$

$$= M^2 N^2 \sum_{\{i:a_i=a\}} \sum_{\{j:b_j=b\}} \left|\int \langle \alpha_i|U_1\varphi\rangle\langle \beta_j|U_2\psi\rangle\langle U_1\varphi \otimes U_2\psi|\Psi\rangle dU_1 dU_2\right|^2 \quad (18)$$

The probability amplitudes (17) defining the joint probability (18) are represented by a linear combination of products of probability amplitudes $\langle \alpha_i|U_1\varphi\rangle$ and $\langle \beta_j|U_2\psi\rangle$. Each probability amplitude $\langle \alpha_i|U_1\varphi\rangle$ corresponds to a transition within system I, from a state $|U_1\varphi\rangle$ to a state defined by an eigenvector $|\alpha_i\rangle$ of $A$ corresponding to the measured eigenvalue $a$, and each probability amplitude $\langle \beta_j|U_2\psi\rangle$ corresponds to a transition within system II, from a state $|U_2\psi\rangle$ to a state defined by an eigenvector $|\beta_j\rangle$ of $B$ corresponding to the measured eigenvalue $b$. The probability amplitudes $\langle \alpha_i|U_1\varphi\rangle$ and $\langle \beta_j|U_2\psi\rangle$ are independent of each other. The coefficients of the linear combination in (18) are the probability amplitudes $\langle U_1\varphi \otimes U_2\psi|\Psi\rangle$. These probability amplitudes determine the correlation between the outcomes of measurements of observables $A$ and $B$. The outcomes of the measurements are correlated due to the initial entanglement, the common cause, without being in a cause-effect relation.

The situation described above is analogous to the situation known in classical probability calculus, where correlation between measurements of two random variables may be present when there is no cause-effect relation between the variables and their measurements. Such correlation may result from an "entanglement" of the initial probability distribution of random variables. In quantum probability calculus, an initial probability distribution is replaced by probability amplitudes. It appears that the mistake Bell and others made trying to interpret outcomes of the measurements performed on mutually non-interacting component systems of a combined system was that they based their arguments on classical probability calculus whereas one should use arguments based on quantum probability theory.

The idea that in the absence of an interaction between component systems and prior to any measurement, the state of the combined system becomes a non-entangled state determined by an earlier interaction between the component systems, was proposed by Bohm and Aharonov [4]: "*after the molecule of spin zero decomposes (…) we suppose that in any individual case, the spin of each atom becomes definite in some direction, while that of the other atom is opposite.*" They noted that their idea was inspired by Einstein's proposal: "*In fact, Einstein has (in a private communication) actually proposed such an idea; namely, that the current formulation of the many-body problem in quantum mechanics may break down when particles are far enough apart.*" However, they further proposed that "*in the statistical sense*" the local collapse leads to "*a uniform probability for any direction.*" This suggests that to obtain the joint probability of outcomes of measurements of spins of two spatially separated atoms, one should use quantum mechanics to calculate the joint probability of outcomes of spin measurements assuming that the state of the combined system is a tensor product of antiparallel spin states of separated atoms corresponding to a given spin direction of the first atom, and then to take an average of the calculated joint probabilities over all directions using classical probability calculus. This approach was proven incorrect by Bell [2] (Eq. (11)).

The local collapse proposal resolves the problems described in the EPR paper. The conclusion that "*as a consequence of two different measurements performed upon the first system, the second system may be left in states with two different wave functions*" is based on an incorrect interpretation of the state of the combined system after the component systems cease to interact. According to the local-collapse postulate presented above, the state of system II before and after the first measurement performed on system I is defined by a state vector $|\zeta\rangle = |U_2\psi\rangle$ resulting from the local collapse of the initially entangled state $|\Psi\rangle$ of the combined system to a non-entangled state defined by the tensor product $|\xi\otimes\zeta\rangle = |U_1\varphi\otimes U_2\psi\rangle$ of vectors in $\mathcal{H}_1$ and $\mathcal{H}_2$ such that $\langle\xi\otimes\zeta|\Psi\rangle \neq 0$. Consequently, the outcome of the first measurement performed on system I does not influence the outcome of the second measurement performed on system II. However, it may be possible that two different states $|\zeta\rangle$ and $|\zeta'\rangle$ corresponding to two different outcomes $|\xi\otimes\zeta\rangle$ and $|\xi'\otimes\zeta'\rangle = |U_1'\varphi\otimes U_2'\psi\rangle$ of the local collapse included in the superposition (13) are eigenstates of two non-commuting observables.

The hypothesis of a global collapse of an entangled state (Eq. (7)) due to a measurement performed on a component system and the postulated local collapse of an initially entangled state when the component systems cease to interact with each other lead to the same formulas for the joint probability (Eq. (8) and (12)) because the superposition (13) is identical with the entangled state $|\Psi\rangle$. However, while the entangled state $|\Psi\rangle$ is the actual state of the combined system comprising two interacting component systems, the superposition (13) is an integral (a continuous superposition) over all non-entangled states which are the possible outcomes of the local collapse of $|\Psi\rangle$ after the component systems cease to interact. The argument in favor of the postulated local collapse is that such interpretation does not require a "*spooky action at a distance*" between the component systems and provides an explanation of the results consistent with quantum mechanics and based on local realism. Such interpretation is also consistent with the description of measurements performed on component systems when the component-system Hamiltonians are non-trivial, especially, when the eigenvalues of the Hamiltonians are non-degenerate. In this case, after the component systems cease to interact with each other, the state of each component system is defined by eigenvectors of its Hamiltonian. The same rule should be true when every state vector of the component system is an eigenvector corresponding to the same eigenvalue – an energy constant. Finally, eq. (12) shows that in the case of multiple measurements performed on component systems, the joint probability of an outcome is independent of how the sequence of measurements performed on one system is merged with the sequence of measurements performed on the other system. Although, strictly speaking, this fact is not a sufficient condition to claim that the measurement performed on one system has no effect on the state of the other system, it would be quite a coincidence if this claim was false. While the view that quantum mechanics is inconsistent with local realism is still widespread, the number of advocates of local realism is growing. In particular, in his recent paper [10] Griffiths proposed to think of the entangled state $|\Psi\rangle$ of the combined system when the component systems cease to interact with each other as a pre-probability. The proposal to treat $|\Psi\rangle$ as a continuous superposition (13) of tensor products of non-entangled states to which the non-interacting component systems collapse provides an argument complementary to Griffiths' claim that "nonlocality claims are consistent with Hilbert space quantum mechanics" by showing that the results used as evidence of nonlocality can be derived within the local realism framework, where the pre-probability $|\Psi\rangle$ determines the probability amplitude $\langle \xi \otimes \zeta | \Psi \rangle$ of the local collapse of the entangled state $|\Psi\rangle$ to a non-entangled state $|\xi \otimes \zeta\rangle$.

## 5. A system of two spin-1/2 particles

The Gedankenexperiment presented in the EPR paper was reformulated by Bohm and Aharonov [4] in terms a system of two spin-1/2 particles in the singlet state. Such system is simpler to describe and more suitable for experimental verification.

For the sake of clarity, known results are included in this section. Let Oxyz be an arbitrary coordinate system and let $x$, $y$, $z$ be the unit vectors of its coordinate axes. Let $\boldsymbol{\sigma} = (\sigma_x, \sigma_y, \sigma_z)$ be the vector of Pauli matrices corresponding to, and hereinafter referred to as spin operators in the Oxyz coordinate system. The eigenvector $|zm\rangle$ of the spin operator $\sigma_z = z \cdot \boldsymbol{\sigma}$ corresponds to the eigenvalue $m$ equal to $1$ or $\bar{1}$.

Let $\boldsymbol{n} \equiv \boldsymbol{n}(\theta, \varphi) = (\sin\theta \cos\varphi, \sin\theta \sin\varphi, \cos\theta)$ be an arbitrary unit vector having spherical coordinates $\theta, \varphi$. Consider the spin operator $\boldsymbol{n} \cdot \boldsymbol{\sigma}$. It is said that $\boldsymbol{n}$ defines the spin direction of the particle. The eigenvector $|nm\rangle$ of the spin operator $\boldsymbol{n} \cdot \boldsymbol{\sigma}$ corresponds to the eigenvalue $m$ equal to $1$ or $\bar{1}$. The eigenvectors $|nm\rangle$ and $|zm\rangle$ are related as follows:

$$|n1\rangle = \cos\frac{\theta}{2}|z1\rangle + \sin\frac{\theta}{2}e^{i\varphi}|z\bar{1}\rangle,$$
$$|n\bar{1}\rangle = -\sin\frac{\theta}{2}e^{-i\varphi}|z1\rangle + \cos\frac{\theta}{2}|z\bar{1}\rangle. \qquad (19)$$

For every unit vector $\boldsymbol{n}$, vectors $|n1\rangle$ and $|n\bar{1}\rangle$ form a basis of the Hilbert space associated with the system comprising one spin-1/2 particle. The parameters used for describing the system consist of the spin direction $\boldsymbol{n}$ and the spin number $m$. The energy of every state of the system can be assumed zero.

Consider a combined system comprising two spin-1/2 particles. Assume that the spin directions of the particles are defined by two arbitrary unit vectors: by $\boldsymbol{a}$ for particle one and by $\boldsymbol{b}$ for particle two. The Hilbert space $\mathcal{H}_1 \otimes \mathcal{H}_2$ of the combined system is the linear span of four basis vectors $|am_1\rangle \otimes |bm_2\rangle$ abbreviated as $|am_1 bm_2\rangle$.

The singlet state of a two spin-1/2 particle system is an entangled state usually defined as

$$|z00\rangle = \frac{1}{\sqrt{2}}(|z1z\bar{1}\rangle - |z\bar{1}z1\rangle). \qquad (20)$$

The quantum numbers in $|zSM\rangle$ denote the total spin numbers $S = 0$ and $M = 0$ of the system. It is easy to verify using Eq. (19) that for every unit vector $\boldsymbol{n}$ one obtains $|\boldsymbol{n}00\rangle = |\boldsymbol{z}00\rangle$.

Let $A = \boldsymbol{a} \cdot \boldsymbol{\sigma_1}$ and $B = \boldsymbol{b} \cdot \boldsymbol{\sigma_2}$ be the two measured observables. Then $A \otimes I_2$ and $I_1 \otimes B$ are observables of the combined system of two spin-1/2 particles system corresponding to $A$ and $B$, respectively. The eigenvectors of $A \otimes I_2$ are $|\boldsymbol{a}m_1\boldsymbol{n}m_2\rangle$ where $\boldsymbol{n}$ can be an arbitrary unit vector. Similarly, the eigenvectors of $I_1 \otimes B$ are $|\boldsymbol{n}m_1\boldsymbol{b}m_2\rangle$ where $\boldsymbol{n}$ can be an arbitrary unit vector. One can now calculate the joint probability $\mathbb{P}(\boldsymbol{a}m_1, \boldsymbol{b}m_2|\boldsymbol{z}00)$ that the measurement of spin of particle one along $\boldsymbol{a}$ gives a value $m_1$ and the measurement of spin of particle two along $\boldsymbol{b}$ gives a value $m_2$, provided that the system was initially in the singlet state $|\boldsymbol{z}00\rangle$ and the particles are far apart and thus do not interact with each other. According to Eq. (8) and (9)

$$\mathbb{P}(\boldsymbol{a}m_1, \boldsymbol{b}m_2|\boldsymbol{z}00) = \mathbb{P}(\boldsymbol{b}m_2|\boldsymbol{a}m_1|\boldsymbol{z}00) = \mathbb{P}(\boldsymbol{a}m_1|\boldsymbol{b}m_2|\boldsymbol{z}00) = |\langle \boldsymbol{a}m_1\boldsymbol{b}m_2|\boldsymbol{z}00\rangle|^2. \quad (21)$$

Replacing $|\langle \boldsymbol{a}m_1\boldsymbol{b}m_2|\boldsymbol{z}00\rangle|^2$ in the joint probability formula (21) with the expression (A8) calculated in Appendix A one obtains

$$\mathbb{P}(\boldsymbol{a}m_1, \boldsymbol{b}m_2|\boldsymbol{z}00) = \frac{1}{4}(1 - \boldsymbol{a} \cdot \boldsymbol{b}\, m_1 m_2) \quad (22)$$

and therefrom one can calculate the quantum correlation defined in Eq. (11):

$$C_q = \sum_{m_1, m_2} m_1 m_2 \mathbb{P}(\boldsymbol{a}m_1, \boldsymbol{b}m_2|\boldsymbol{z}00) = -\boldsymbol{a} \cdot \boldsymbol{b}. \quad (23)$$

Quantum correlation (23) was first calculated by Bell [2]. It violates Bell-type inequalities, for example, the CHSH inequality proposed in [5]. Experiments performed on pairs of spin-1/2 particles but mostly on pairs of polarization-entangled photons also show that Bell-type inequalities are violated [13]. However, violation of Bell-type inequalities does not imply that the outcomes of the two spin measurements are in a cause-effect relation. The probability amplitude $\langle \boldsymbol{a}m_1\boldsymbol{b}m_2|\boldsymbol{z}00\rangle$ defining the joint probability (21) can be written as (Eq. (17))

$\langle am_1 bm_2|z00\rangle$

$$= 4\int \langle am_1|U_1|n_1 m_1'\rangle\langle bm_2|U_2|n_2 m_2'\rangle\langle n_1 m_1'\ n_2 m_2'|U_1\otimes U_2|z00\rangle dU_1 dU_2 \quad (24)$$

or, using spherical coordinates defining the spin directions $\boldsymbol{n}_1 = \boldsymbol{n}(\theta_1, \varphi_1)$ and $\mathbf{n}_2 = \boldsymbol{n}(\theta_2, \varphi_2)$,

$\langle am_1 bm_2|z00\rangle$

$$= \frac{1}{4\pi^2}\int_0^\pi \sin(\theta_1)\, d\theta_1 \int_0^{2\pi} d\varphi_1 \int_0^\pi \sin(\theta_2)\, d\theta_2 \int_0^{2\pi} d\varphi_2 \langle am_1|\boldsymbol{n}(\theta_1,\varphi_1)m_1'\rangle\langle bm_2|\boldsymbol{n}(\theta_2,\varphi_2)m_2'\rangle$$

$$\times \langle \boldsymbol{n}(\theta_1,\varphi_1)m_1'\ \boldsymbol{n}(\theta_2,\varphi_2)m_2'|z00\rangle. \quad (25)$$

Equation (25) is to be understood that after the two particles cease to interact with each other, they locally collapse to a non-entangled state $|\boldsymbol{n}(\theta_1,\varphi_1)m_1'\ \boldsymbol{n}(\theta_2,\varphi_2)m_2'\rangle$, but it is impossible to know neither the spin directions $\boldsymbol{n}_i$ nor the spin values $m_i'$ of this state. Only the joint probability $\mathbb{P}(\boldsymbol{a}m_1, \boldsymbol{b}m_2|z00)$ of the measurement outcomes is known. The probability amplitudes $\langle \boldsymbol{n}(\theta_1,\varphi_1)m_1'\ \boldsymbol{n}(\theta_2,\varphi_2)m_2'|z00\rangle$ of the collapse present in the sum (25) define the probability amplitude $\langle \boldsymbol{a}m_1 \boldsymbol{b}m_2|z00\rangle$ and the joint probability $\mathbb{P}(\boldsymbol{a}m_1, \boldsymbol{b}m_2|z00)$. They "remember" the singlet state of the two particles prior to the collapse to a non-entangled state and are responsible for the correlation between the outcomes of the two measurements.

**Acknowledgements**

The author wishes to emphasize the role of the Internet in his work and to thank all those who provide valuable, freely accessible contributions to its contents. Special thanks are also due to my beloved wife Wiesia for her support and enthusiasm about my work. Finally, the author expresses his gratitude to his employer, Signify, which encourages its employees to pursue their interests by supporting a healthy balance between work and personal life.

**Appendix A**

For convenience, the derivation of the transition probabilities $|\langle \boldsymbol{a}m\boldsymbol{b}m|\boldsymbol{z}00\rangle|^2$ is outlined below using the notation introduced in section 5. Let $\theta_a, \varphi_a$ and $\theta_b, \varphi_b$ be the spherical coordinates of the respective unit vectors $\boldsymbol{a}$ and $\boldsymbol{b}$, in the Oxyz coordinate system. Using Eq. (19) one can express vectors $|\boldsymbol{a}m_1\boldsymbol{b}m_2\rangle$ as linear combinations of vectors $|\boldsymbol{z}m_1\boldsymbol{z}m_2\rangle$:

$$|\boldsymbol{a}1\boldsymbol{b}1\rangle = \cos\frac{\theta_a}{2}\cos\frac{\theta_b}{2}|z1z1\rangle + \cos\frac{\theta_a}{2}\sin\frac{\theta_b}{2}e^{i\varphi_b}|z1z\bar{1}\rangle$$

$$+\sin\frac{\theta_a}{2}\cos\frac{\theta_b}{2}e^{i\varphi_a}|z\bar{1}z1\rangle + \sin\frac{\theta_a}{2}\sin\frac{\theta_b}{2}e^{i(\varphi_a+\varphi_b)}|z\bar{1}z\bar{1}\rangle, \qquad (A1)$$

$$|\boldsymbol{a}1\boldsymbol{b}\bar{1}\rangle = -\cos\frac{\theta_a}{2}\sin\frac{\theta_b}{2}e^{-i\varphi_b}|z1z1\rangle + \cos\frac{\theta_a}{2}\cos\frac{\theta_b}{2}|z1z\bar{1}\rangle$$

$$-\sin\frac{\theta_a}{2}\sin\frac{\theta_b}{2}e^{i(\varphi_a-\varphi_b)}|z\bar{1}z1\rangle + \sin\frac{\theta_a}{2}\cos\frac{\theta_b}{2}e^{i\varphi_a}|z\bar{1}z\bar{1}\rangle \qquad (A2)$$

$$|\boldsymbol{a}\bar{1}\boldsymbol{b}1\rangle = -\sin\frac{\theta_a}{2}\cos\frac{\theta_b}{2}e^{-i\varphi_a}|z1z1\rangle - \sin\frac{\theta_a}{2}\sin\frac{\theta_b}{2}e^{-i(\varphi_a-\varphi_b)}|z1z\bar{1}\rangle$$

$$+\cos\frac{\theta_a}{2}\cos\frac{\theta_b}{2}|z\bar{1}z1\rangle + \cos\frac{\theta_a}{2}\sin\frac{\theta_b}{2}e^{i\varphi_b}|z\bar{1}z\bar{1}\rangle, \qquad (A3)$$

$$|\boldsymbol{a}\bar{1}\boldsymbol{b}\bar{1}\rangle = \sin\frac{\theta_a}{2}\sin\frac{\theta_b}{2}e^{-i(\varphi_a+\varphi_b)}|z1z1\rangle - \sin\frac{\theta_a}{2}\cos\frac{\theta_b}{2}e^{-i\varphi_a}|z1z\bar{1}\rangle$$

$$- \cos\frac{\theta_a}{2}\sin\frac{\theta_b}{2}e^{-i\varphi_b}|z\bar{1}z1\rangle + \cos\frac{\theta_a}{2}\cos\frac{\theta_b}{2}|z\bar{1}z\bar{1}\rangle. \qquad (A4)$$

Using the definition of the singlet state given in Eq. (20) one obtains:

$$|\langle \boldsymbol{a}m\boldsymbol{b}m|\boldsymbol{z}00\rangle|^2 = \frac{1}{2}\left(\cos^2\frac{\theta_a}{2}\sin^2\frac{\theta_b}{2} + \sin^2\frac{\theta_a}{2}\cos^2\frac{\theta_b}{2} - \frac{1}{2}\sin\theta_a\sin\theta_b\cos(\varphi_a - \varphi_b)\right)$$

$$= \frac{1}{4}(1 - \cos\theta_a\cos\theta_b - \sin\theta_a\sin\theta_b\cos(\varphi_a - \varphi_b)) = \frac{1}{4}(1 - \boldsymbol{a}\cdot\boldsymbol{b}), \qquad (A6)$$

$$|\langle a m b \bar{m} | z 0 0 \rangle|^2 = \frac{1}{2}\left(\cos^2\frac{\theta_a}{2}\cos^2\frac{\theta_b}{2} + \sin^2\frac{\theta_a}{2}\sin^2\frac{\theta_b}{2} + \frac{1}{2}\sin\theta_a\sin\theta_b\cos(\varphi_a - \varphi_b)\right)$$

$$= \frac{1}{4}(1 + \cos\theta_a\cos\theta_b + \sin\theta_a\sin\theta_b\cos(\varphi_a - \varphi_b)) = \frac{1}{4}(1 + \boldsymbol{a}\cdot\boldsymbol{b}). \quad (A7)$$

Hence,

$$|\langle a m_1 b m_2 | z 0 0 \rangle|^2 = \frac{1}{4}(1 - \boldsymbol{a}\cdot\boldsymbol{b}\, m_1 m_2) \quad (A8)$$